\newif\ifsingle

\newif\ifproofs

\ifsingle
\documentclass[11pt,draftclsnofoot, onecolumn]{IEEEtran}		
\else		
\documentclass[10pt,final, twocolumn]{IEEEtran}
\fi

\usepackage{times}
\usepackage{amsmath,dsfont}
\usepackage{amssymb,amsthm}
\usepackage{epsfig,verbatim}
\usepackage{subfigure}
\usepackage{setspace}
\usepackage{color}
\usepackage{xcolor}
\usepackage{cite}
\usepackage{epstopdf}
\usepackage{graphics}
\usepackage{graphicx}
\usepackage{accents}
\usepackage{acronym}
\usepackage[bookmarks,colorlinks]{hyperref}
\usepackage{booktabs}
\usepackage{mathtools}
\usepackage{afterpage}
\usepackage{float}
\usepackage[linesnumbered,ruled,vlined]{algorithm2e}
\usepackage{algpseudocode}
\usepackage{enumitem}
\usepackage{bm}
\usepackage{lipsum, nccmath}

\usepackage{cuted}
\DeclareMathOperator*{\argmin}{arg\,min}

\newtheorem{lemma}{Lemma}

\newcommand{\removelatexerror}{\let\@latex@error\@gobble}
\newcommand{\myVec}[1]{{\boldsymbol{#1}}}
\newcommand{\myMat}[1]{{\boldsymbol{#1}}}
\newcommand{\mySet}[1]{\mathcal{#1}}

\newcommand{\Kiter}{K} 

\acrodef{bc}[BC]{broadcast channel}
\acrodef{mac}[MAC]{multiple access channel}  
\acrodef{adc}[ADC]{analog-to-digital convertor}  
\acrodef{csi}[CSI]{channel state information} 
\acrodef{snr}[SNR]{signal-to-noise ratio}
\acrodef{sinr}[SINR]{signal-to-interference-and-noise ratio}
\acrodef{tx}[TX]{Transmitter}  
\acrodef{mimo}[MIMO]{multiple-input multiple-output}
\acrodef{mse}[MSE]{mean-squared error}
\acrodef{pdf}[PDF]{probability density function}
\acrodef{rv}[RV]{random variable} 
\acrodef{isi}[ISI]{intersymbol interference}  
\acrodef{awgn}[AWGN]{additive white Gaussian noise} 
\acrodef{lti}[LTI]{linear time-invariant}  
\acrodef{ut}[UT]{user terminal} 
\acrodef{mmw}[mmWave]{millimeter wave}
\acrodef{dma}[DMA]{dynamic metasurface antenna}
\acrodef{ofdm}[OFDM]{orthogonal frequency division multiplexing}
\acrodef{dnn}[DNN]{deep neural network}
\acrodef{gnn}[GNN]{graph neural network}
\acrodef{ml}[ML]{machine learning}
\acrodef{dl}[DL]{deep learning}
\acrodef{mlp}[MLP]{Multilayer perceptron}
\acrodef{sgd}[SGD]{stochastic gradient descent}
\acrodef{bpsk}[BPSK]{binary phase shift keying}
\acrodef{pgd}[PGD]{projected gradient descent}
\acrodef{ber}[BER]{bit error rate}
\acrodef{csi}[CSI]{channel state information}
\acrodef{map}[MAP]{maximum a-posteriori probability}
\acrodef{ct}[CT]{continuous-time}
\acrodef{manet}[MANET]{mobile ad hoc network}
\acrodef{noma}[NOMA]{non-orthogonal multiple access}
\acrodef{sic}[SIC]{successive interference cancellation}
\acrodef{mmse}[MMSE]{minimum mean squared error}
\acrodef{lmmse}[LMMSE]{linear minimum mean-squared error}

\definecolor{blue}{rgb}{0,0,1}

\ifsingle

\else

\fi 

\IEEEoverridecommandlockouts

\title{Rapid Optimization of Superposition Codes for Multi-Hop NOMA MANETs via Deep Unfolding
}
\author{
	\IEEEauthorblockN{Tomer Alter,~\IEEEmembership{Student~Member,~IEEE,} and Nir Shlezinger,~\IEEEmembership{Senior~Member,~IEEE,}\\
	} 
	\thanks{
    Parts of this work were presented in the IEEE Military Communications Conference (MILCOM 2023) as the paper~\cite{alter2023deep}.
	}
 \IEEEauthorblockA{
                     School of ECE, Ben-Gurion University of the Negev, Israel. \{tomeralt@post.bgu.ac.il; nirshl@bgu.ac.il\}
}

}
\allowdisplaybreaks
\begin{document}

\maketitle
\pagestyle{plain}
\thispagestyle{plain}
\begin{abstract}
Various communication technologies are expected to utilize \acp{manet}. By combining \acp{manet} with \ac{noma} communications, one can support scalable, spectrally efficient, and flexible network topologies. To achieve these benefits of \ac{noma} \acp{manet}, one should determine the transmission protocol, particularly the superposition code. However, the latter involves lengthy optimization that has to be repeated when the topology changes. 
In this work, we propose an algorithm for rapidly optimizing superposition codes in multi-hop \ac{noma} \acp{manet}. To achieve reliable tunning with few iterations, we adopt the emerging {\em deep unfolding} methodology, leveraging data to boost reliable settings. Our superposition coding optimization algorithm utilizes a small number of projected gradient steps while learning its per-user hyperparameters to maximize the minimal rate over past channels in an unsupervised manner. The learned optimizer is designed for both settings with full \acl{csi}, as well as when the channel coefficients are to be estimated from pilots. We show that the combination of principled optimization and machine learning yields a scalable optimizer, that once trained, can be applied to different topologies. We cope with the non-convex nature of the optimization problem by applying parallel-learned optimization with different starting points as a form of ensemble learning. Our numerical results demonstrate that the proposed method enables the rapid setting of high-rate superposition codes for various channels.

\end{abstract}

\acresetall
\section{Introduction}

    A broad range of communication technologies, ranging from industrial systems~\cite{wollschlaeger2017future} to drone swarms~\cite{chen2020toward} and tactical networks~\cite{cha2018robust},  are subject to strict demands in terms of connectivity, robustness, mobility, and security. An emerging paradigm shift aimed at meeting these requirements is to deviate from the uplink/downlink star topology operation of conventional wireless communications, e.g., cellular networks. Among the leading approaches for realizing dynamic and flexible networks is the \ac{manet} technology \cite{burbank2006key}, where communication links are established during operation, possibly on demand and over multi-hop routes. Forming multi-hop communication topologies on demand allows mobile communicating entities to collaborate in communication and decision-making~\cite {shlezinger2022collaborative}. 
	
	One of the core challenges in multi-user communications, which grows more prominent in  \acp{manet} compared to traditional topologies, is the presence and treatment of cross interference. Such interference naturally arises from the fact that multiple users share the same temporal and spectral channel resources. Arguably the most common strategy in conventional wireless communications to deal with cross-interference is to mitigate it, i.e.,  to boost 
	 orthogonality among the communicating entities. This can be achieved via the division of the spectral and temporal resources, or alternatively via media access protocols designed to prevent collisions. 
 While orthgonalizing multi-user communications facilitates system design, it comes at the cost of throughput degradation due to its sub-optimal usage of the channel resources~\cite{el2011network}. Consequently, recent years have witnessed a growing interest in transiting from orthogonality-based communications to \ac{noma} solutions \cite{Liu2017SurveyNOMA, Ding2017SurveyNOMA}. \ac{noma}  allows users to simultaneously share the channel resources for supporting heterogeneous end-devices. However, non-orthogonal multi-user communications inevitably impose interference.
	
	Various receive methods have been proposed for enabling reliable communications in the presence of interference \cite{andrews2005interference}. A widely adopted receiver algorithm for non-orthogonal \acp{bc} is \ac{sic}, which is based on successively decoding a set of superimposed messages that are transmitted to different users over a shared channel. The gains of \ac{sic} are also indicated in its ability to approach the achievable rate region of such channels when combined with {\em superposition coding}~\cite{Liu2017SurveyNOMA, Ding2017SurveyNOMA,loung2021deep}, whilst its complexity only grows linearly with the number of users. Similar approaches for handling interference via superposition coding and interference cancellation at the receiver were also proposed for \acp{mac}~\cite{gao2017theoretical,shlezinger2019deepSIC}. 

    The theoretical benefits of \ac{noma} are widely studied in the context of conventional uplink/downlink wireless communications, based on \ac{bc}/\ac{mac} modeling, respectively. While these studies consider star topologies, recent works have identified that these gains can also be harnessed in multi-hop topologies~\cite{jain2020performance}, and thus potentially also in \acp{manet}. In order to harness these gains of \ac{noma}, one should determine the superposition code utilized by all communicating entities based on the current channel realization. In \acp{manet}, which are inherently designed for dynamic and mobile settings, the channel realizations, and even the network topology, often vary rapidly. Thus, superposition code setting has to be repeated frequently based on possibly noisy channel estimates. Nonetheless, this procedure involves lengthy optimization even for conventional topologies, with recent studies proposing iterative methods~\cite{amer2023resource} and the usage of machine learning architectures~\cite{chowdhury2021ml,schynol2023coordinated}. However, the setting of superposition codes is notably more complicated in \acp{manet}, particularly when using noisy channel estimates. This motivates the design of efficient optimization techniques for rapidly and reliably configuring \ac{noma} \acp{manet}. 


    In this work, we propose an algorithm that rapidly optimizes superposition code for multi-hop \ac{noma} \acp{manet}. Our proposed method, based on unfolding \ac{pgd} optimization, is coined {\em Unfolded PGDNet}.   Unfolded PGDNet tunes the superposition code for a given \ac{manet} based on noisy limited pilots, using estimated \ac{csi} as indicative features. We derive our method by formulating the superposition code design as a non-convex constrained optimization problem aiming at maximizing the minimal rate in the network. We then leverage emerging learn-to-optimize methodologies~\cite{chen2022learning,shlezinger2023model}, and particularly deep unfolding~\cite{shlezinger2020model,shlezinger2022model}, to learn from data how to enable projected gradient steps to rapidly recover a suitable configuration and cope with noisy channel estimates. 

    As the minimal rate objective of the superposition coding optimization is non-convex, we tackle it by proposing an ensemble method~\cite{sagi2018ensemble}. In particular, we apply in parallel multiple instances of the learned optimizer with different initial settings; the algorithm then identifies multiple candidate solutions and selects the one with the best minimal rate. While we train Unfolded PGDNet for a given topology, we show that its interpretable operation renders it transferable to multiple different topologies. Accordingly, Unfolded PGDNet facilitates rapid adaptation of superposition codes in dynamic \acp{manet}, coping not only with variations in the channel realizations for a given topology, but also with variations in the topology itself.  Our numerical results demonstrate that Unfolded PGDNet rapidly produces solid superposition code configurations, reducing the latency compared with conventional optimization by factors varying from $10\times$ to over $80\times$, as well as consistently outperforming competing data-driven techniques based on \acp{gnn}~\cite{shen2022graph}.

The rest of this work is organized as follows: Section~\ref{sec:System} describes the system model; the proposed learn-to-optimize algorithm for configuring \ac{noma} \acp{manet} is detailed in Section~\ref{sec:Unfolded}, and is numerically evaluated in Section~\ref{sec:Sims}. Finally, Section~\ref{sec:Conclusions} provides concluding remarks.

  	Throughout the paper, we use boldface lower-case letters for vectors, e.g., ${\myVec{x}}$;
	the $i$th element of ${\myVec{x}}$ is written as $[{\myVec{x}}]_i$.  
	Matrices are denoted with boldface upper-case letters,  e.g., 	$\myMat{M}$;   $[\myMat{M}]_{i,j}$   is its $(i,j)$th element, while $[\myMat{M}]_{i,:}$ is its $i$th row. 
	Calligraphic letters, such as $\mySet{X}$, are used for sets, while $\mathbb{R}$ denotes the real numbers. 
	The $\ell_2$ norm, transpose, and Hermitian transpose  are written as $\| \cdot \|$, $(\cdot)^T$, and $(\cdot)^H$,   respectively.


\section{System Model}
\label{sec:System}


In this section, we formulate the system model for multi-hop \ac{noma} \acp{manet}. We commence with presenting the communication system in Subsection~\ref{subsec:ComModel}, and how \ac{csi} is acquired in Subsection~\ref{subsec:Estimation}. Based on these, we formulate the superposition coding optimization problem in Subsection~\ref{subsec:Problem}, which constitutes the starting point for our derivation of Unfolded PGDNet in the sequel. 

\subsection{Communication System Model} 
\label{subsec:ComModel}
\begin{figure}
    \centering
    \includegraphics[width=0.8\columnwidth]{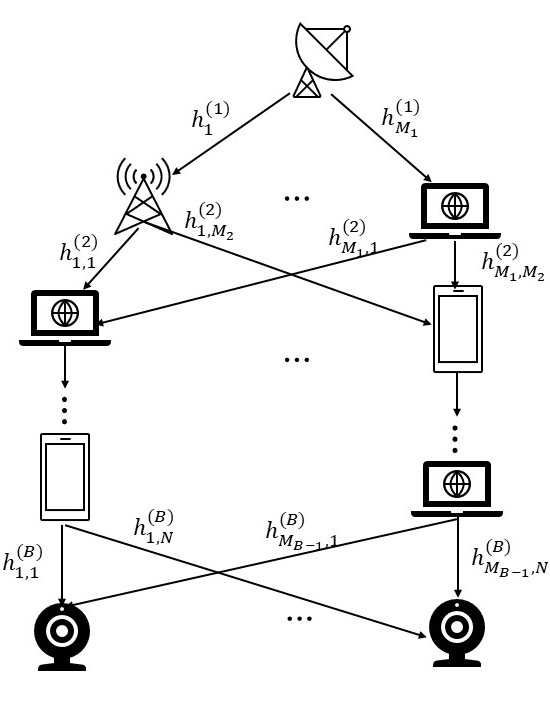}
   \vspace{-0.2cm}
    \caption{Multi-hop \ac{manet} illustration.} 
    \label{fig:Model1}
\end{figure}

We consider a block-fading multi-hop scalar \ac{manet} with a single transmitter and multiple receiver nodes  (end-users). There are no direct links between the transmitter and the receivers, and communication is achieved only via intermediate relays spreading over $B$ hops. 
The channels and the \ac{manet} topology  are block-wise time-varying, representing the ad-hoc nature of such networks. In each coherence duration of index $t$, the transmitter wishes to communicate with $N(t)$ end-users, while  the $b$th hop involves $M_b(t)$ relays for each $b\in \{1,\ldots,B\}\triangleq\mySet{B}$, with $M_B(t)\equiv N(t)$, as the $B$th hop reaches the end-users.
An example topology is illustrated in Fig.~\ref{fig:Model1}.

The communicating devices employ {\em power domain \ac{noma}}~\cite{Ding2017SurveyNOMA}. Accordingly, letting $s_n(t)$ denote the signal intended for the $n$th end-user at the $t$th coherence duration, $n \in \{1,\ldots,N(t)\}\triangleq \mySet{N}(t)$, the transmitter generates a linear combination of $\{s_n(t)\}_{n\in\mySet{N}(t)}$ via {\em superposition coding} weighted by the non-negative power coefficients $\{\varphi_n(t)\}_{n\in\mySet{N}(t)}$. Let $h_{m}^{(1)}(t)$  denote the channel coefficient between the transmitter and the $m$th relay of the first hop, which is constant during the coherence duration by the block-fading model \cite[Ch. 3]{tse2005fundamentals}. The signal received by the $m$th relay of the first hop is given by
\begin{align} 
     y_{m}^{(1)}(t)& = h_{m}^{(1)}(t)\sum_{n=1}^{N(t)} \varphi_n(t) s_n(t) + w_{m}^{(1)}(t) ,   
     \label{eq:bc received message}
\end{align}
for each $m\in \{1,\ldots,M_1(t)\}$. In \eqref{eq:bc received message},
 $w_{m}^{(b)}$ is \ac{awgn} with variance $\sigma_b^2$ for each hop  $b \in \mySet{B}$. 

The relays operate in a {\em decode-and-forward} manner~\cite[Ch. 16]{el2011network}, aiming to recover $\{s_n(t)\}_{n\in \mySet{N}(t)}$ from their corresponding channel output~\eqref{eq:bc received message}. Then, $\{s_n(t)\}_{n\in \mySet{N}(t)}$  are relayed to the next hop, with each relay employing a dedicated superposition code. We use $\{p_{m,n}^{(b)}(t)\}_{n\in\mySet{N}(t)}$ to denote the power coefficients of the $m$th relay at the $b$th hop. Let $h_{m, i}^{(b)}(t)$ be the channel from the $m$th relay to the $i$th receiving node at the $b$th hop over the $t$th coherence period. The signal received at this receiving node is obtained as 
\begin{align} 
    y_{i}^{(b)}(t)& = \sum_{m=1}^{M_b(t)}h_{m,i}^{(b)}(t) \sum_{n=1}^{N(t)}p_{m,n}^{(b)}(t)s_n(t) + w_i^{(b)}(t),  
    \label{eq:ma received message}
\end{align}
for each $i \in \{1,\ldots,M_b(t)\}$. For clarity, we summarize the variables defined above in Table~\ref{tab:variables}.

Decoding at both the relays and the end-users is carried out using \ac{sic}, whose combination with superposition coding is known to approach the capacity region of single-hop \ac{noma} systems~\cite{Ding2017SurveyNOMA}. 
\ac{sic} operates iteratively: first the signal with the highest \ac{sinr} is decoded, while treating all the other signals as noise. Then, it is subtracted from the superimposed message and the signal with the second largest \ac{sinr} is recovered in the same fashion. This procedure is carried out over all $N(t)$ messages at the relays (which decode all messages for decode-and-forward), while the $i$th end-user terminates \ac{sic} when decoding $s_i(t)$.

\begin{table} 
    \centering
    \caption{List of Variables and Parameters}
    \label{tab:variables}
    \begin{tabular}{ll}
        \hline
        \textbf{Symbol} & \textbf{Definition} \\
        \hline
        $w_{i}^{b}(t)$ & i.i.d AWGN noise at $i$th receiver, $b$th hop \\
        $h_{i}^{(1)}(t)$ & transmitter - $m$th relay at first hop channel  \\
        $h_{m,i}^{(b)}(t)$ & $m$th relay - $i$th receiver at $b$th hop channel \\
        $s_n(t)$ & signal intended for $n$th end-user  \\
        $\varphi_n(t)$ & transmit power allocate to $n$th signal at first hop \\
        $p_{m,n}^{(b)}(t)$ & $b$th hop $m$th relay power allocate to $n$th signal \\
        \hline
    \end{tabular}
\end{table}

\subsection{Channel State Information}
\label{subsec:Estimation}
The block-fading nature of the considered channel model indicates that the communicating entities are likely not to have \ac{csi}, i.e., prior knowledge of the channel coefficients. Hence, they would be required to estimate it from periodic pilots. To that aim, we assume that each coherence duration is commenced with a transmission of $T$ pilots, whose quantity is always larger than the number of transmitters in each hop, i.e., $T \geq \max_{b,t}M_b(t)$. Let $\myVec{u}_i^{(b)}$ denote the $T\times 1$ pilots vector sent by the $i$th transmitter at the $b$th hop. Following conventional pilot modeling, e.g.,~\cite{hassibi2003much,shlezinger2018spectral}, it is assumed that the pilots are orthonormal, i.e., that  $\big(\myVec{u}_i^{(b)}\big)^H\myVec{u}_l^{(b)} = 1$ if $l=i$ and zero otherwise.

A candidate approach to utilize the pilots to estimate the channel uses the  \ac{lmmse} estimator. To formulate this estimator, let $\myVec{y}_{i}^{(b)}(t)$ be the $T\times 1$ vector representing the channel outputs corresponding to the pilots obtained at the $i$th receiving node of the $b$th hop during the $t$th block. Assuming that the channel coefficients are all zero-mean and with variance $\sigma_h^2$, the  \ac{lmmse} estimate of the channel $h_m^{(1)}(t)$ is~\cite[Ch. 8]{Papoulis2002probability}
\begin{align}
\label{eq:BC LMMSE}
   &\hat{h}_m^{(1)}(t) = \frac{\sigma_h^2}{\sigma_h^2 + \sigma_1^2}\big(\myVec{u}_1^{(1)}\big)^H\myVec{y}_m^{(1)}(t).
\end{align}
%
Similarly, at subsequent hops, $h_{m, i}^{(b)}(t)$ is estimated as
\begin{align}
\label{eq:MAC LMMSE}
    &\hat{h}_{m,i}^{(b)}(t)= \frac{\sigma_h^2}{\sigma_h^2 + \sigma_b^2}\big(\myVec{u}_m^{(b)}\big)^H\myVec{y}_i^{(b)}(t).
\end{align}

\subsection{Problem Formulation}
\label{subsec:Problem}
\subsubsection{Achievable Rates}
The communication model for the considered \ac{noma} \ac{manet} is formulated in Subsection~\ref{subsec:ComModel} for an arbitrary superposition code, dictated by the power coefficients $\{\varphi_n(t)\}$ and $\{p_{m,n}^{(b)}(t)\}$. However, the exact setting of the code coefficients notably affects the achievable communication rate in the \ac{manet}. In particular, by \eqref{eq:bc received message}  the $n$th message is recoverable by the $m$th relay of the first hop (with arbitrarily small error probability) if it is coded with a rate (in bits per channel use), not larger than~\cite[Ch. 6]{tse2005fundamentals}
\begin{align} 
        R_{m,n}^{{(1)}}(t) &= \log_2\left(1 + \frac{|h_{m}^{(1)}(t)|^2 \varphi_n(t)^2}{|h_{m}^{(1)}(t)|^2 \sum\limits_{\varphi_i(t) \in \Phi_n(t)} \varphi_i(t)^2 + \sigma_{1}^2}\right),
    \label{eqn: bc rate}
\end{align}
where we defined
\begin{equation*}
\Phi_n(t) \triangleq \{\varphi_i(t): i\in \mySet{N}(t) \backslash \{n\}, \varphi_i(t) \leq \varphi_n(t) \}.    
\end{equation*}

The message also has to be recoverable by each relay as well as its corresponding end-user. In particular, the $l$th message is decoded by all relays, as well as by the $l$th end-user, by recovering all messages preceding it via \ac{sic}. By \eqref{eq:ma received message}, the $n$th message can be recovered by the $l$the receiving node at the $b$th hop when its rate is not larger than 
\begin{align} 
        R_{l,n}^{{(b)}}(t) &=  \log_2\left(1 + \frac{g_{l,n}^{(b)}(t)}{\sum\limits_{ g_{l,i}^{(b)}(t) \in  G_{l,n}^{(b)}(t)}  g_{l,i}^{(b)}(t) + \sigma_{b}^2}\right),
    \label{eqn: ma rate}
\end{align}
for all hop $b\in \mySet{B}  \backslash \{1\}$. In \eqref{eqn: ma rate},  we defined:
\begin{align} 
      g_{i, n}^{(b)}(t) &\triangleq \left|\sum_{m=1}^{M}h_{m,i}^{(b)}(t) p_{m,n}^{(b)}(t)\right|^2, \qquad n,i\in\mySet{N}(t) ,
    \label{eqn: ma gain}
\end{align}
and 
\begin{equation*}
G_{l,n}^{(b)}(t) \triangleq \left\{g_{l,i}^{(b)}(t): i \in \mySet{N}(t) \backslash \{n\}, g_{l,i}^{(b)}(t) \leq g_{l,n}^{(b)}(t)\right\}.
\end{equation*}
\subsubsection{Superposition Code Optimization}
The rate expressions in \eqref{eqn: bc rate} and \eqref{eqn: ma rate} depend on the superposition coding coefficients, written henceforth as the $\bar{M}(t) \times N(t)$ matrix $\myMat{P}(t)$, where $\bar{M}(t) \triangleq 1+\sum_{b=1}^{B-1}M_b(t)$. 
The matrix $\myMat{P}(t)$ is comprised of $B-1$ sub-matrices $\myMat {P}^{(b)}(t)\in\mathbb{R}^{M_b(t)\times N(t)}, b=1,2,\dots,B-1$, and a vector $\myVec{\varphi}(t)\in\mathbb{R}^{N(t)}$. The entries of the sub-matrices are  given by $[\myMat {P}^{(b)}(t)]_{m,n} = p_{m,n}^{(b)}(t)$ for $m \in \mySet{M}_b(t)$ where $\mySet{M}_b(t)\triangleq \{1,2,\dots, M_b(t)\}$, while $\myVec {\varphi}(t) = [\varphi_1(t),\varphi_2(t),\dots,\varphi_{N(t)}(t)]$. 
To create $\myMat{P}(t)$, we stack  
\begin{equation}
    \myMat{P}(t)\triangleq\left[
    \begin{matrix}
        \myMat {P}^{(1)}(t)\\
        \vdots \\
        \myMat {P}^{(B-1)}(t)\\
        \myVec {\varphi}^T(t)
    \end{matrix}
    \right].
    \label{eqn:PDef}
\end{equation}   

Using these notations combined with \eqref{eqn: ma rate} and \eqref{eqn: bc rate}, and by stacking the \ac{manet} channel coefficients as the vector $\myVec{h}(t)$,  the rate supported for the $n$th message is given by
\begin{align}
    R_n(\myMat{P}(t),\myVec{h}(t)) = \min\Big\{&\min_{l\in\mySet{N}(t): g_{l,n}^{(B)}(t) \geq g_{l,l}^{(B)}(t)} R_{l, n}^{(B)}(t), 
    \notag \\
    &\min_{m\in\mySet{M}_b(t), b\in \mySet{B}\backslash\{B\}} R_{m,n}^{(b)}(t) \Big\}.
    \label{eqn:RnP}
\end{align}
In \eqref{eqn:RnP}, the first minimization term corresponds to the fact that each end-user does not have to recover all messages, only those preceding and including its own in the superposition code. However, each relay decodes all messages (as encapsulated in the second term), and thus the coding rate must be set such that the message can be reliably decoded by all these nodes. 

We aim to set the superposition coding coefficients $\myMat{P}(t)$ to maximize the {\em minimal rate}, i.e., the throughput one can guarantee for all end-users, being a key performance measure in \acp{manet}. Without loss of generality, we assume a unit power constraint on each of the transmitting devices, i.e., $\myMat{P}(t) \in \mySet{P}(t)$, where $\mySet{P}(t)$ is defined as
\begin{equation}
\label{eqn:PSet}
    \mySet{P}(t) = \left\{\myMat{P}\in[0,1]^{\bar{M}(t) \times N(t)}: \big\|[\myMat{P}]_{m,:}\big\|_2= 1; \forall m \right\}.
\end{equation}
Accordingly, our goal is to map the received pilots into a superposition code, i.e., knowledge of 
\begin{equation}
 \mySet{Y}(t) \triangleq \big\{ \big\{ \myVec{y}_{m}^{(b)}(t)\big\}_{m\in \mySet{M}_b(t)}\big\}_{b\in \mySet{B}},   \label{eqn:YsetDef}
\end{equation} 
and $\myVec{\sigma} = \{\sigma_{b}\}_{b=1}^{B}$, into the coefficients matrix
\begin{align} 
        \myMat{P}^{\star}(t) &= \mathop{\arg\max}\limits_{\myMat{P}(t)\in \mySet{P}(t)} \min_{n\in\mySet{N}(t)}  R_n(\myMat{P}(t),\myVec{h}(t)).
    \label{eqn: best power allocation}
\end{align}
\subsubsection{Challenges}
The superposition code design problem formulated in \eqref{eqn: best power allocation} gives rise to several challenges. These challenges that are associated with the statement of the problem in  \eqref{eqn: best power allocation}, as well as with the dynamic nature of \acp{manet} and its block-faded model, as summarized below:
\begin{enumerate}[label={C\arabic*}]
    \item \label{itm:nonCvx} The optimization problem is non-convex, thus recovering the optimal $ \myMat{P}^{\star}(t)$ may be infeasible. Moreover, even finding a good setting of $\myMat{P}(t)$ (in the sense of increasing the minimal rate) is likely to involve lengthy iterative solvers.
    \item \label{itm:rapid} The superposition code has to be set anew each time the channel coefficients changes. Thus, the design has to be carried out rapidly, e.g., within a small and fixed number of iterative steps.
    \item \label{itm:noisy csi} The channel realizations, which are necessary for rate calculation, are not available. Instead, they must estimated from noisy pilots, resulting in inaccurate \ac{csi}. 
    \item \label{itm:time varying} One of the main properties of \acp{manet} is their dynamic topology, which changes over time. A solver to \eqref{eqn: best power allocation} must be flexible thus to different topologies.
\end{enumerate}

To facilitate tackling \eqref{eqn: best power allocation} while coping with \ref{itm:nonCvx}-\ref{itm:time varying}, we assume access to  a set of $T_r$ past  channel realizations. This data is available offline, i.e., when designing the solver, and can be obtained from past measurements. The resulting data set is denoted by
\begin{equation}
\label{eqn:DataSet}
    \mySet{D} \triangleq \{\myVec{h}(t), \myVec{\sigma}(t)\}_{t=-T_r}^{-1},
\end{equation}
with $\myVec{h}(t)$ denoting the stacking of the \ac{manet} channel realizations at block $t$, while $\myVec{\sigma}(t)$ stacks the noise variances. 
The availability of data indicates the possibility of leveraging machine learning tools, combined with principled optimization to cope with \ref{itm:nonCvx}-\ref{itm:time varying}, as proposed in the following section.

\section{Unfolded Superposition Coding Optimization}
\label{sec:Unfolded}
This section presents our proposed data-aided optimizer for  rapid superposition code design. We gradually derive our method based on the following steps: 
\begin{itemize}
    \item We first formulate the \ac{pgd} steps for tackling \eqref{eqn: best power allocation} in Subsection~\ref{subsec:PGD}.  \ac{pgd} is suitable for convex problems~\cite[Ch. 9]{boyd2004convex}; it assumes \ac{csi};  and it typically requires many iterations to converge. Accordingly, it is not tailored to tackle \ref{itm:nonCvx}-\ref{itm:time varying} on its own. Instead, it  is used as our basis optimizer due to its simplicity and interpretability.
    \item Then, in Subsection~\ref{subsec:PGDNet} we convert \ac{pgd} into PGDNet, which is designed to tackle \ref{itm:nonCvx}-\ref{itm:rapid} (without dealing with \ref{itm:noisy csi}-\ref{itm:time varying} yet).
    We assume full \ac{csi} and a static \ac{manet} topology. For such setting, we leverage data to optimize the \ac{pgd} optimizer, i.e.,  we learn-to-optimize rapidly~\cite{lavi2023learn}, via deep unfolding 
methodology~\cite{shlezinger2020model,shlezinger2022model} (thus tackling \ref{itm:rapid}), and extend it to cope with \ref{itm:nonCvx} using an ensemble of Unfolded PGDNet model. 
    \item Next, we deal with the fact that instead of \ac{csi}, one has noisy pilots (\ref{itm:noisy csi}). To cope with this, we extend our design in Subsection~\ref{subsec:estData} to process \ac{lmmse} features, and adapt its learning procedure accordingly.
    \item  The last step shows that the usage of machine learning to augment the \ac{pgd} optimizer enables operation in  time-varying topologies, thus coping with~\ref{itm:time varying}, as  described in Subsection~\ref{subsec:blockFad}.  
\end{itemize} 
We next elaborate on these steps, followed by a dedicated in discussion in Subsection~\ref{subsec: Discussion}.

\subsection{Projected Gradient Descent for NOMA MANETs}
\label{subsec:PGD}

Problem \eqref{eqn: best power allocation} represents constrained maximization with $N(t)\cdot \bar{M}(t)$  optimization variables. While the problem is non-convex, one can identify a plausible setting for the superposition code parameters $\myMat{P}(t)$ by applying iterative \ac{pgd} steps starting from some suitable initialization $\myMat{P}^{(0)}(t)$.   
The resulting iterative procedure at iteration index $k$ is given by
\begin{align} 
    \myMat{P}^{(k+1)}(t)  = &\Pi_{\mySet{P}(t)}\Big(\myMat{P}^{(k)}(t) \notag \\
    &+ \mu^{(k)} \nabla_{\myMat{P}(t)} \min_{n\in\mySet{N}(t)} R_n\big( \myMat{P}^{(k)}(t), \myVec{h}(t)\big)\Big), 
    \label{eqn:PGDStep}
\end{align}
where $\mu^{(k)}$ is the step-size, and $ \Pi_{\mySet{P}(t)}(\cdot)$ denotes the projection operator onto $\mySet{P}(t)$. As the projection operator onto the simplex \eqref{eqn:PSet} typically involves additional iterative procedures~\cite{wang2013projection}, we approximate it using the following operator (which guarantees that the projection of any matrix with at least one positive entry per column indeed lies in $\mySet{P}(t)$ by \eqref{eqn:PSet}), written as  $\tilde{\myMat{P}}(t) = \hat{\Pi}_{\mySet{P}(t)}(\myMat{P}(t))$, with
\begin{equation}
    [\tilde{\myMat{P}}(t)]_{m,:} = \frac{1}{ \big\|[\myMat{P}(t)]_{m,:}^+\big\|_2} [\myMat{P}(t)]_{m,:}^+.
    \label{eqn:Proj}
\end{equation}
In \eqref{eqn:Proj},  $(\cdot)^+$ is the positive part operator, i.e., $a^+ \triangleq \max(0,a)$, applied element-wise.

To implement \ac{pgd}, the gradients of the objective function are required. These are stated in the following lemma.
\begin{lemma}
\label{lem:RateGrad}
For a given $n\in\mySet{N}(t)$, let $\tilde{m}\in\{1,2,\dots,\max_i M_i(t)\}$, $\tilde{b}\in\mySet{B}\backslash\{B\}$, and $l_{B}\in\mySet{N}(t)$, be the indices holding $\tilde{m},\tilde{b} = \argmin_{m,b}R_{m,n}^{(b)}$ and $l_{B} = \argmin_{l}R_{l,n}^{(B)}$.
Then, for each $q \in\mySet{N}(t)$ and $s \in\mySet{M}_b(t)$ it holds that    
\begin{subequations}     
\label{eqn:RateGrad}  
\begin{align}
    \frac{\partial R_n(t)}{\partial \varphi_q(t)} &= 
    \begin{cases}
        \frac{\partial R^{(\tilde{b})}_{\tilde{m},n}(t)}{\partial \varphi_q(t)} & \tilde{b}=1; R_{\tilde{m},n}^{(\tilde{b})}(t) < R_{l_{B},n}^{(B)}(t) \\
        0 & {\rm else}
    \end{cases}\\
    \frac{\partial R_n(t)}{\partial  p_{s,q}^{(b)}(t)} &= 
    \begin{cases}
        \frac{\partial R^{(\tilde{b})}_{\tilde{m},n}(t)}{\partial p_{s,q}^{(\tilde{b})}(t)} & b=\tilde{b}\neq 1,B;  R_{\tilde{m},n}^{(\tilde{b})}(t) < R_{l_{B},n}^{(B)}(t) \\
        \frac{\partial{R_{l_{B},n}^{(B)}(t)}}{\partial{p_{s,q}^{(B)}(t)}} & b=B; R_{\tilde{m},n}^{(\tilde{b})}(t) \geq R_{l_{B},n}^{(B)}(t) \\
         0 & {\rm else}
    \end{cases}
\end{align}
\end{subequations}
    where the gradients of $ R^{(1)}_{m,n}(t),R^{(b)}_{l,n}(t)$ are given in \eqref{eqn:RateGrad2}.
    \begin{figure*}
    \begin{subequations}      
    \label{eqn:RateGrad2} 
     \begin{align}
     \frac{\partial R^{(1)}_{m,n}(t)}{\partial \varphi_q(t)} &=\frac{1}{\rm ln2} \begin{cases}
     0 & \varphi_q(t) > \varphi_n(t)\\
     \frac{2|h_m^{(1)}(t)|^2 \varphi_n(t)}{|h_m^{(1)}(t)|^2 \sum\limits_{i \in \mySet{N}(t)/n: \varphi_i(t) > \varphi_n(t) }{(\varphi_i(t))}^2 + \sigma_{1}^2} & \varphi_q(t) = \varphi_n(t)\\
     \frac{-2|h_m^{(1)}(t)|^4(\varphi_n(t))^2\varphi_q(t)}{|h_m^{(1)}(t)|^2\sum\limits_{i \in \mySet{N}(t)/n: \varphi_i(t) \geq \varphi_n(t) }(\varphi_i(t))^2 + \sigma_{1}^2} \frac{1}{|h_m^{(1)}(t)|^2\sum\limits_{i \in \mySet{N}(t)/n: \varphi_i(t) > \varphi_n(t) }(\varphi_i(t))^2 + \sigma_{1}^2} & \varphi_q(t) < \varphi_n(t)
     \end{cases} \label{eqn:RateGradRe} \\
      \frac{\partial R^{(b)}_{l,n}(t)}{\partial p_{s,q}^{(b)}(t)} &=\frac{1}{\rm ln2} \begin{cases}
     0 & g_{l,q}^{(b)}(t) > g_{l,n}^{(b)}(t)\\
     \frac{2|h_{l,s}^{(b)}(t)|^2 p_{s,q}^{(b)}(t)}{{\sum\limits_{i \in \mySet{N}(t)/n: g_{l,i}^{(b)}(t) >g_{l,n}^{(b)}(t) }g_{l,i}^{(b)}(t) + \sigma_{b}^2}} & g_{l,q}^{(b)}(t) = g_{l,n}^{(b)}(t)\\
     \frac{-2|h_{l,s}^{(b)}(t)|^2p_{s,q}^{(b)}(t)g_{l,n}^{(b)}(t)}{\sum\limits_{i \in \mySet{N}(t)/n: g_{l,i}^{(b)}(t) \geq g_{l,n}^{(b)}(t) }g_{l,i}^{(b)}(t) + \sigma_{b}^2}\frac{1}{\sum\limits_{i \in \mySet{N}(t)/n: g_{l,i}^{(b)}(t) >g_{l,n}^{(b)}(t) }g_{l,i}^{(b)}(t) + \sigma_{b}^2} & g_{l,q}^{(b)}(t) < g_{l,n}^{(b)}(t) 
     \label{eqn:RateGradRx}
     \end{cases}
     \end{align}     
    \end{subequations}
     \end{figure*}
    \end{lemma}

\begin{IEEEproof}
    See Appendix~\ref{app:Proof1}.
\end{IEEEproof}

The resulting optimization method with $K$ iterations is summarized as Algorithm~\ref{alg:PGD} (where the time index $t$ is omitted). Note that the computation requires knowledge of the \ac{csi}, denoted by the vector $\myVec{h}(t)$ which includes all channel realizations in the \ac{manet}. Moreover, \ac{pgd} is inherently designed for convex objectives, and typically requires $K$ to be large to achieve convergence with a fixed step-size~\cite{parikh2014proximal}. While this implies that the \ac{pgd} optimizer in Algorithm~\ref{alg:PGD} is on its own not suitable for the problem at hand, we use it as a form of {\em inductive bias} to obtain a data-aided optimizer capable of coping with  \ref{itm:nonCvx}-\ref{itm:time varying} in the following subsections.

  \begin{algorithm}
    \caption{ $\textbf{PGD}_{\Kiter}(\cdot; {\myMat{P}^{(0)}},\myVec{\mu}) $}
    \label{alg:PGD} 
    \SetKwInOut{Initialization}{Init}
    \Initialization{Iterations $\Kiter$;
      Step-sizes $\myVec{\mu}=\{\mu^{(k)}\}_{k=0}^{K-1}$; \\ Initial guess ${\myMat{P}^{(0)}}$ }
    \SetKwInOut{Input}{Input}
    \Input{\ac{csi} $\myVec{h}, \myVec{\sigma} $}  
    {
        \For{$k = 0, 1, \ldots \Kiter$ }{%
                    Calculate $\nabla_{\myMat{P}} \min_{n\in\mySet{N}(t)} R_n\big( \myMat{P}, \myVec{h}\big) $ using \eqref{eqn:RateGrad}\;
                    Update $\myMat{P}^{(k+1)}$ via \eqref{eqn:PGDStep} \label{line:P_Wa}\;
                }
    \KwRet{$\myMat{P}^{(K)}$}

  }
\end{algorithm}

\subsection{Unfolded PGDNet}
\label{subsec:PGDNet}
Let us first assume a static topology of the \ac{manet} and access to full \ac{csi}, thus ignoring \ref{itm:noisy csi}-\ref{itm:time varying} for the time being. Accordingly, in this subsection (as well as in Subsection~\ref{subsec:estData}) we omit the block index $t$ for brevity. 

Algorithm \ref{alg:PGD} optimizes the power allocation for a given \ac{manet} realization. However, its convergence speed highly depends on the hyperparameters, i.e., the step-sizes $\myVec{\mu}$, and the resulting rate can be notably affected by the initial guess $\myMat{P}^{(0)}$ due to the non-convexity. 
We thus propose a data-aided implementation of \ac{pgd}, termed {\em Unfolded PGDNet}. We derive our method based on the following steps: $(i)$ We first tackle \ref{itm:rapid}  converting \ac{pgd} into a fixed-latency discriminative trainable architecture~\cite{shlezinger2022discriminative} via deep unfolding methodology~\cite{shlezinger2022model}, and propose dedicated training algorithm; $(ii)$  Then, we extend it into an ensemble of multiple parallel models that converge to different local optima to cope with \ref{itm:nonCvx}.

\subsubsection{PGDNet Architecture}
We propose to leverage data to {\em learn-to-optimize}  based on deep unfolding~\cite{shlezinger2022model}. This methodology converts iterative optimizers with a predetermined number of iterations into machine learning models. 
The proposed Unfolded PGDNet applies \ac{pgd} (Algorithm~\ref{alg:PGD}) with a predefined (and small) number of iterations $\emph{K}$. By doing that, we guarantee an a-priori known, and considerably low, run-time. While the accuracy of first-order optimizers such as \ac{pgd} is usually invariant of hyperparameters setting when allowed to run until convergence (as long as the step-sizes are sufficiently small)~\cite{boyd2004convex}, its performance is largely affected by these settings when the number of iterations is fixed.

\begin{figure*}
    \centering
    \includegraphics[width=0.8\linewidth]{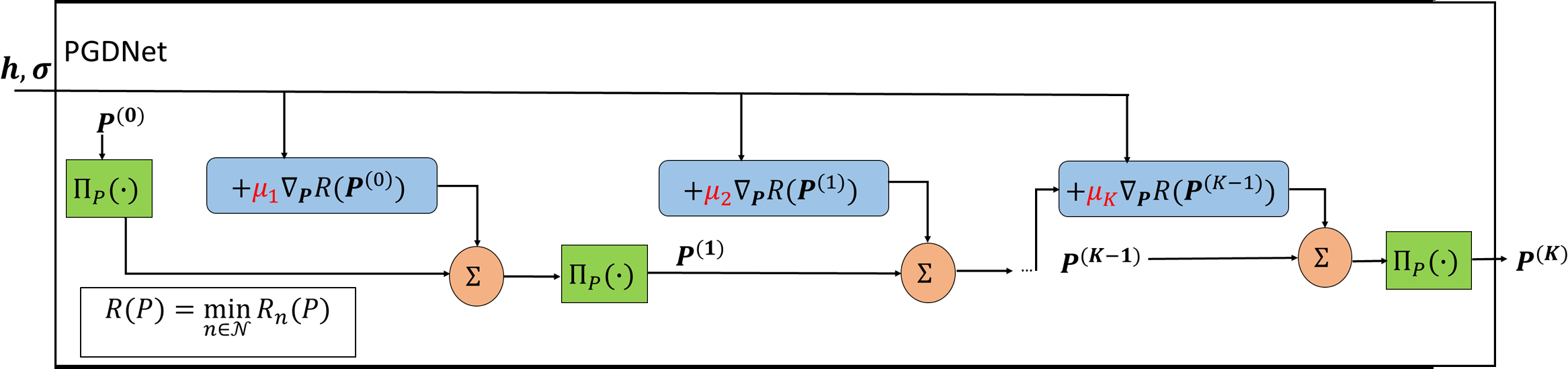}
   \vspace{-0.2 cm}
    \caption
    { PGDNet architecture; trainable parameters are marked in \textcolor{red}{red}. } 
    \label{fig:Net Architecture}
\end{figure*}

In Unfolded PGDNet, we treat the hyperparameters of the iterative optimizer, i.e., $\myVec{\mu}$, as the {\em trainable parameters of a machine learning model}. The resulting architecture can be viewed as a form of a \ac{dnn} with $K$ layers: Each layer of index $k$ implements a single \ac{pgd} iteration, and has a single trainable parameter, which is the step-size $\mu_k$. The resulting architecture is illustrated in Fig.~\ref{fig:Net Architecture}. As in Algorithm~\ref{alg:PGD}, we denote its operation  for \ac{csi} $\myVec{h}, \myVec{\sigma}$ with hyperparameters $\myVec{\mu}$ and initial guess $\myMat{P}^{(0)}$ as $\textbf{PGD}_{\Kiter}(\myVec{h}, \myVec{\sigma} ; \myMat{P}^{(0)}, \myVec{\mu})$.

\subsubsection{Training PGDNet}
The training procedure aims to set the hyperparameters $\myVec{\mu}$ to make PGDNet maximize the minimal rate for the channel available in the data set $\mySet{D}$ in \eqref{eqn:DataSet}. As in~\cite{lavi2023learn}, we also exploit the interpretability of the trainable architecture to facilitate and regularize the training procedure. Specifically, we account for the fact that the features exchanged between layers correspond to superposition coding coefficients, i.e., $\myMat{P}^{(k)}$ for the $k$th layer, and encourage the model to learn gradually improved settings. 

To formulate the resulting learning objective, we write $\textbf{PGD}_{\Kiter}^{(k)}(\cdot ; \cdot)$ as the output of the $k$th iteration of PGDNet, i.e., $\myMat{P}^{(k)}$. We train PGDNet by setting the loss function to be
\begin{align}
 \mathcal{L}_{\mathcal{D}}(\myVec{\mu}) = &\frac{-1}{|\mathcal{D}|} \sum_{(\myVec{h},\myVec{\sigma})\in \mathcal{D}} \sum_{k=1}^K\log_{2}(1+k) \notag \\ &\times\min_{n\in\mySet{N}(t)} R_n\left(\textbf{PGD}_{\Kiter}^{(k)}(\myVec{h}, \myVec{\sigma} ; \myMat{P}^{(0)}, \myVec{\mu}), \myVec{h} \right),   
 \label{eqn:trainLoss}
\end{align}
where the logarithmic weights are used to have a growing contribution to the overall loss of the more advanced learned iterations~\cite{samuel2019learning}.
Note that \eqref{eqn:trainLoss} enables {\em unsupervised learning}, i.e., there is no need to provide `ground truth` power allocation. This stems from the task being an optimization problem, for which one can evaluate each setting of the optimization variables. 

The learn-to-optimize method, summarized as Algorithm~\ref{alg:PGDNet}, uses data to tune $\myVec{\mu}$ based on \eqref{eqn:trainLoss}.  
We initialize $\myVec{\mu}$ before the training process with a fixed step-size with which PGD converges (given sufficient iterations). Then, we exploit the differentiability of the gradients in \eqref{eqn:RateGrad} and the projection operator in \eqref{eqn:Proj} to tune $\myVec{\mu}$ via conventional deep learning tools (where in Algorithm~\ref{alg:PGDNet} we employ mini-batch stochastic gradient descent for learning). 
We modify the initial guess $\myMat{P}^{(0)}$ during training to have the learned hyperparameters be suitable for different starting points.
At the end of the training process, the learned vector $\bm{\mu}$ is used as a hyperparameter for  rapid finding of the best power allocation for a given channel via $K$ iterations of Algorithm~\ref{alg:PGD}.

\begin{algorithm}
    \caption{Training PGDNet}
    \label{alg:PGDNet}
    \SetKwInOut{Initialization}{Init}
    \Initialization{Set $\bm{\mu}$ as fixed step-sizes; \\
        Set learning rate $\eta$, batches $Q$, and epochs.}
     \SetKwInOut{Input}{Input}
    \Input{
    Training set  $\mathcal{D}$}
  
    {
        \For{$i = 1, 2, \ldots, {\rm epochs}$}{%
                    Randomly divide  $\mathcal{D}$ into $Q$ batches $\{\mathcal{D}_q\}_{q=1}^Q$
                    
                    \For{$q = 1, \ldots, Q$}{
                    Set random initial guess  ${\myMat{P}^{(0)}}$ \;
                    \label{stp:loss}
                     Compute  average loss $\mathcal{L}_{\mathcal{D}_q}(\bm{\mu})$  using \eqref{eqn:trainLoss}\;                    
                    Update  $\bm{\mu}\leftarrow \bm{\mu} - \eta \nabla_{\bm{\mu}} \mathcal{L}_{\mathcal{D}_q}(\bm{\mu})$
                    }
                 
                    }
        \KwRet{$\bm{\mu}$}
  }
\end{algorithm}

\begin{figure}[htbp]
    \centering
    \subfigure[Unfolded PGDNet training]{\includegraphics[width=0.6\columnwidth]{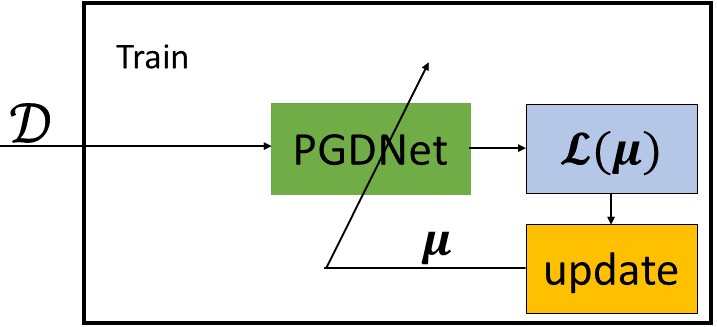}}
    \hfill
    \subfigure[Unfolded PGDNet ensemble]{\includegraphics[width=0.7\columnwidth]{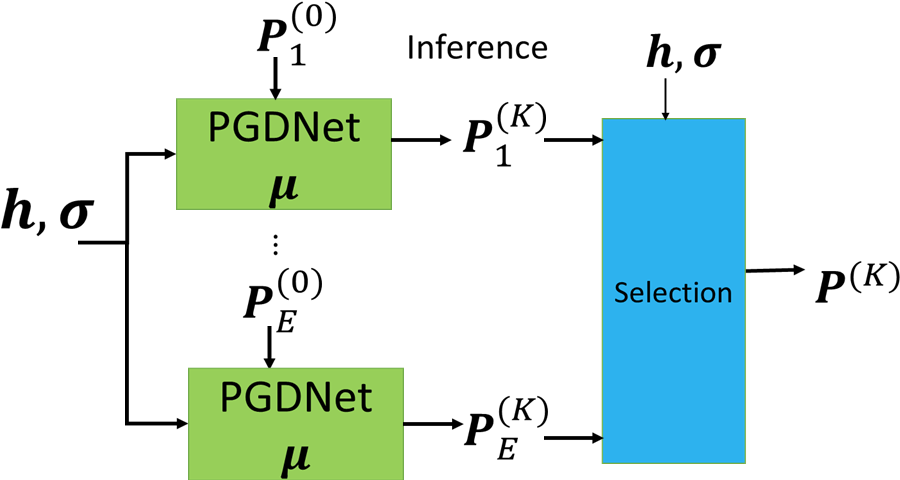}}
    \caption{Unfolded PGDNet training and inference procedures}
    \label{fig:training and inference}
\end{figure}

\subsubsection{PGDNet Ensemble}
Unfolded PGDNet uses data to make \ac{pgd} operate most efficiently within $K$ iterations (where $K$ is small). However, by \ref{itm:nonCvx} it is still expected to converge to a power allocation in the proximity of the initial guess $\myMat{P}^{(0)}$, which is not necessarily the most suitable one. Nonetheless, by setting $K$ to be relatively small, one can apply Unfolded PGDNet $E > 1$ times, each time with a different $\myMat{P}^{(0)}$ (thus accounting for \ref{itm:nonCvx}), while still inferring sufficiently fast to cope with \ref{itm:rapid}.

The resulting power allocation procedure thus utilizes an ensemble model~\cite{sagi2018ensemble} comprised of a parallel application of $E$ Unfolded PGDNet algorithms. Specifically, we train a single Unfolded PGDNet model, and learn the hyperparameters $\myVec{\mu}$. Then, in runtime, given a channel realization $\myVec{h},\myVec{\sigma}$, we apply Unfolded PGDNet with these hyperparameters $E$, using different initial guesses denoted $\myMat{P}^{(0)}_1,\ldots,\myMat{P}^{(0)}_E$, as illustrated in Fig.~\ref{fig:training and inference}(b). To select the superposition code, we  leverage the fact that we can evaluate each suggested setting using the optimization objective in \eqref{eqn: best power allocation}, and select the power allocation that gives the maximal min-rate among all the iterations of the ensemble models, i.e., 
\begin{equation}
    \max \limits_{e\in 1,\ldots E } \min_{n\in\mySet{N}(t)}  R_n\left(\textbf{PGD}_{\Kiter}(\myVec{h}, \myVec{\sigma} ; \myMat{P}^{(0)}_e, \myVec{\mu}), \myVec{h}\right).
    \label{eqn:EnsembleOutput}
\end{equation}

\subsection{Learned Optimization from Noisy Pilots}
\label{subsec:estData}
Unfolded PGDNet detailed in the previous Subsection leverages data to enable \ac{pgd} optimization of superposition codes while coping with \ref{itm:nonCvx}-\ref{itm:rapid}. However, it still requires the \ac{csi} to be provided to compute the optimization objective, as in the conventional \ac{pgd}, while in our setting one has access only to noisy pilots by \ref{itm:noisy csi}. To tackle this, we next extend Unfolded PGDNet to optimize from noisy pilots by leveraging \ac{lmmse} \ac{csi} estimates as input features.

\subsubsection{Inference}
To carry out inference, i.e., set a superposition code, using Unfolded PGDNet, we use the \ac{lmmse} estimator formulated in Subsection~\ref{subsec:Estimation} to translate the noisy pilots in \eqref{eqn:YsetDef} into {\em noisy \ac{csi} features}. Then, Unfolded PGDNet is applied as if these features represent the full \ac{csi}. The resulting inference procedure is summarized as Algorithm~\ref{alg:NoisyPGDNet}.

\begin{algorithm}
    \caption{PGDNet with Noisy Pilots}
    \label{alg:NoisyPGDNet}
    \SetKwInOut{Initialization}{Init}
    \Initialization{Trained step-sizes $\bm{\mu}$; Initial $\{\myMat{P}_e^{(0)}\}$; \\
                    Noise and channel variances $\myVec{\sigma}, \sigma_h^2$.}
     \SetKwInOut{Input}{Input}
    \Input{Noisy pilots $\mySet{Y}$}  
    {
        Compute $\hat{\myVec{h}}$ from $\mySet{Y}$ via \eqref{eq:BC LMMSE} and \eqref{eq:MAC LMMSE}\;
        \For{$e = 1, 2, \ldots, E$}{%
                    Compute $\myMat{P}_e = \textbf{PGD}_{\Kiter}\big(\hat{\myVec{h}}, \myVec{\sigma} ; \myMat{P}^{(0)}_e, \myVec{\mu}\big)$
                 
                    }
        \KwRet{$\myMat{P} = \mathop{\arg \max}\limits_{\{\myMat{P}_e\}}  \mathop{\min}\limits_{n\in\mySet{N}(t)}  R_n(\myMat{P}_e,\hat{\myVec{h}})$}
  }
\end{algorithm}

\subsubsection{Training}
The inference procedure in Algorithm~\ref{alg:NoisyPGDNet} applies PGDNet while treating the noisy \ac{csi} estimates as the real \ac{csi}. While this can in general lead to performance degradation, we train PGDNet to produce learned hyperparameters that result in optimized rates concerning the {\em true \ac{csi}} available in the data \eqref{eqn:DataSet}.

\begin{figure}
    \centering
    \includegraphics[width=\columnwidth]{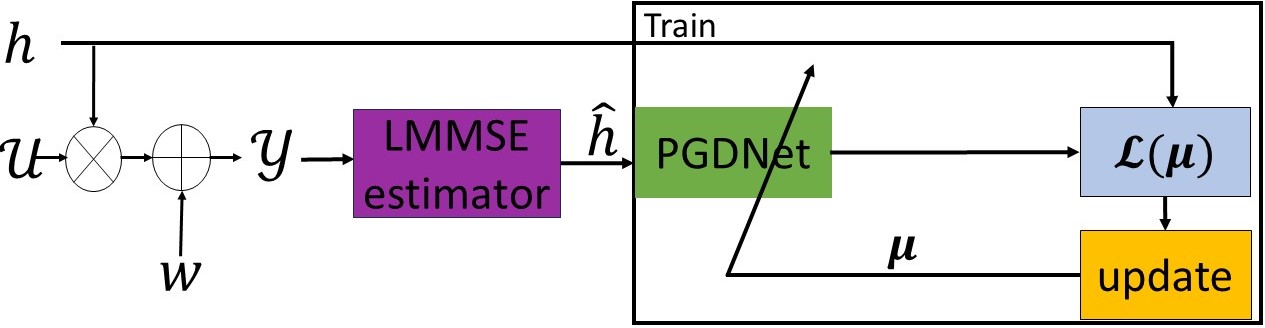}
   \vspace{-0.2cm}
    \caption
    { Training Unfolded PGDNet with noisy pilots}
    \label{fig:noisy training}
\end{figure}

Consequently, the training procedure is based on Algorithm~\ref{alg:PGDNet}, with the core exception that the input is the \ac{lmmse} features, i.e., {\em noisy \ac{csi} estimates}, while the loss function in \eqref{eqn:trainLoss} is computed with the {\em true \ac{csi}}. Specifically, to compute the loss in Step~\ref{stp:loss} of Algorithm~\ref{alg:PGDNet}, we first use the observed channels in $\mySet{D}$ to simulate random channel observations via~\eqref{eq:bc received message}-\eqref{eq:ma received message}, from which the \ac{lmmse} estimate $\hat{\myVec{h}}$ is computed via \eqref{eq:BC LMMSE}-\eqref{eq:MAC LMMSE}. Then, the loss gradients are taken not based on \eqref{eqn:trainLoss}, but rather from 
\begin{align}
 \mathcal{L}_{\mathcal{D}}(\myVec{\mu}) = &\frac{-1}{|\mathcal{D}|} \sum_{(\myVec{h},\myVec{\sigma})\in \mathcal{D}} \sum_{k=1}^K\log_{2}(1+k) \notag \\ &\times\min_{n\in\mySet{N}(t)} R_n\left(\textbf{PGD}_{\Kiter}^{(k)}(\hat{\myVec{h}}, \myVec{\sigma} ; \myMat{P}^{(0)}, \myVec{\mu}), {\myVec{h}} \right),   
 \label{eqn:trainLossNoisy}
\end{align}
Namely, Unfolded PGDNet is applied to the batch of estimated \ac{csi} realizations, while the loss is computed using the true \ac{csi}. 
The procedure is illustrated in Fig.~\ref{fig:noisy training}.

\subsection{Coping with Block-Fading Topologies}
\label{subsec:blockFad}
Our design of Unfolded PGDNet so far considers a given topology. However, as noted in~\ref{itm:time varying}, a key feature of \acp{manet} is their dynamic nature, i.e., some links may be removed from the network and other devices may join the network after its setup, both ending up changing the topology.
Additionally, even if the topology remains the same as we started with, due to the block-fading communication model, some links can be changed over time.
While our design so far does not explicitly deal with these temporal variations, we advocate that its interpretable architecture, i.e., the fact that what is learned are hyperparameters of \ac{pgd} optimization, enable it to cope with~\ref{itm:time varying}.

In particular, we note that conventional \ac{pgd} in Algorithm~\ref{alg:PGD} can be applied to different topologies. Moreover, its hyperparameters $\myVec{\mu}$, which are learned and optimized from data via Algorithm~\ref{alg:PGDNet}, are invariant of the \ac{manet} topology. Consequently, PGDNet trained on a given topology is in fact {\em transferable} to alternative topologies, as we numerically demonstrate in Section~\ref{sec:Sims}.  This allows a set of learned hyperparameters $\myVec{\mu}$ learned from past channel realizations via Algorithm~\ref{alg:PGDNet} with the loss in \eqref{eqn:trainLossNoisy} to be used by PGDNet inference in Algorithm~\ref{alg:NoisyPGDNet} for block-fading \acp{manet}.


\subsection{Discussion}
\label{subsec: Discussion}
Unfolded PGDNet is designed to produce superposition codes for a given \ac{noma} \ac{manet} while coping with the non-convexity of the problem (\ref{itm:nonCvx}), the need to design rapidly (\ref{itm:rapid}), and the noisy and dynamic nature of \acp{manet} (\ref{itm:noisy csi}-\ref{itm:time varying}).  The former is tackled by employing an ensemble of convex optimizers and selecting the best candidate setting computed; the rapid operation of each optimizer is achieved by using data to optimize the optimizer via deep unfolding; and coping with dynamic settings follows from a dedicated learning procedure combined with the flexibility of its hybrid model-based/data-driven design. While classic approaches to improve convergence in terms of the number of iterations, e.g., line search~\cite[Ch. 9]{boyd2004convex}, come at the cost of increased latency due to additional processing carried out at each iteration, our unfolded method does not induce any excessive latency during inference compared to conventional \ac{pgd} with fixed pre-determined step-sizes. 
As opposed to the application of machine learning architectures based on conventional \acp{dnn}, our data-aided algorithm is highly flexible to variations in the \ac{manet} topology.
This makes our approach highly suitable for facilitating the rapid tuning of superposition codes.

The formulation of Unfolded PGDNet is invariant of the initial guesses $\myMat{P}^{(0)}_1(t),\ldots,\myMat{P}^{(0)}_E(t)$. In our numerical study reported in Section~\ref{sec:Sims} we use equal power allocation for $\myMat{P}^{(0)}_1(t)$ and randomize the remaining guesses uniformly over $\mySet{P}(t)$. This setting was empirically shown to be beneficial in terms of rapidly finding high min-rate superposition codes. Nonetheless, one can explore different settings of the initial guesses, which we leave for future investigation.
Moreover, our derivation copes with noisy \ac{csi} by using \ac{lmmse} features, which require knowledge of the channel variance $\sigma_h^2$. One can replace these with a learned feature extractor which does not require such knowledge and can be possibly trained jointly with PGDNet. 
Finally, we note that we consider a centralized setting of the superposition code, i.e., all the superposition coding coefficients are computed jointly. A candidate extension of our methodology would replace this centralized optimization with a decentralized one such that each device can set its superposition coding coefficients following a set of exchanged messages, leveraging recent advances in combining deep unfolding with such distributed optimization, e.g.,~\cite{noah2023limited}.
We leave these extensions of our proposed algorithm for future investigation.

\section{Numerical Evaluations}
\label{sec:Sims}
In this section, we numerically evaluate Unfolded PGDNet\footnote{The source code used in our empirical study along with the hyperparameters is available at \url{https://github.com/AlterTomer/Deep-Unfolded-PGD}}. 
Unfolded PGDNet and the considered benchmarks, whose configuration is detailed in Subsection~\ref{subsec: experimental setups}, are evaluated in setting with gradually increased complexity. 
We start with static \ac{manet}s, where we first assume access to full \ac{csi} data to evaluate coping with \ref{itm:nonCvx}-\ref{itm:rapid} (Subsection~\ref{subsec: static manets}), and proceed to using estimated channels (Subsection~\ref{subsec: static manets noisy}), thus accounting for \ref{itm:noisy csi}. We  conclude by considering dynamic \ac{manet}s, where \ref{itm:nonCvx}-\ref{itm:time varying} are coped with, in Subsection~\ref{subsec:time varying nets}.


\subsection{Experimental Setups}
\label{subsec: experimental setups}
\subsubsection{\ac{manet} Setups}
We simulate \acp{manet} with $B=2$ hops. We consider two different topologies: $(i)$ a $1\times 2\times 2$ \ac{manet} with $M_2(t)=N(t)=2$ end-users  and $M_1(t)=2$ relays;
$(ii)$  a $1\times 3\times 3$ \ac{manet} with $M_2(t)=N(t)=3$ end-users  and $M_1(t)=3$ relays. The channels obey Rayleigh fading, i.e., the channel gains are randomized from a unit variance Gaussian distribution, with  $|\mySet{D}|=1000$ realizations used for learning, and $200$ realizations for  test.

\subsubsection{Optimization Methods}
We employ Unfolded PGDNet with $K=40$ iterations, trained over $100$ epochs using Adam~\cite{kingma2014adam}. We use uniform allocation for the initial guess used in training, i.e., $[\myMat{P}^{(0)}_1(t)]_{m,n}=\frac{1}{\sqrt{N}}$.
We compare our optimizer with the following benchmarks:
\begin{itemize}
    \item {\em Classic \ac{pgd}} -- Algorithm~\ref{alg:PGD}  with fixed step-sizes (manually tuned to achieve consistent convergence). 
    \item {\em \ac{gnn}} -- A data-driven \ac{gnn} based on the architecture proposed in~\cite{shen2022graph}. The \ac{gnn} input is a weighted undirected graph signal representing the channels for each link, and its output is the power assigned to each node in the graph. We set the \ac{gnn} to have the same architecture presented in~\cite{shen2022graph}, which includes two convolutional layers followed by two linear layers. The \ac{gnn} is trained with the same data and objective as Unfolded PGDNet, i.e.,  to maximize the minimal rate. 
    \item  {\em Grid Capacity} --  In the $1\times2\times2$ case, where the considered setup has only $(N(t)-1) (1+M_1(t)) = 3$ optimization variables, we also compute the maximal min-rate by an exhaustive search over all allocations in the unit cube with a grid of resolution of $10^{-2}$. The outcome of this exhaustive search, which is not feasible for larger \acp{manet}, serves as an upper bound on the min-rate of the considered optimizers.
\end{itemize}
\subsection{Static MANETs with Full \ac{csi}}
\label{subsec: static manets}
\begin{figure}
    \centering
    \includegraphics[width=\columnwidth]{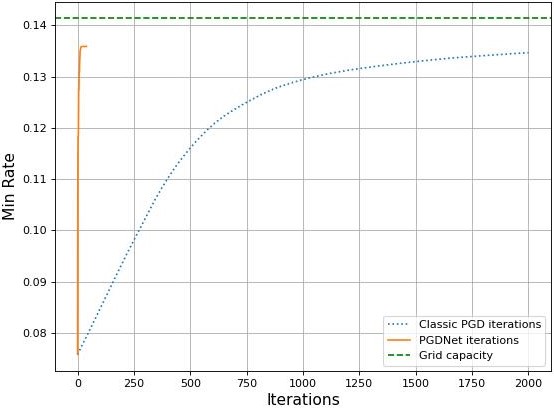}
   \vspace{-0.2cm}
    \caption{Min-rate per PGD iteration averaged over $200$ channels, $1\times 2\times 2$ static \ac{manet}.} 
    \label{fig:PGDNet vs PGD iters M=N=2}
\end{figure}

We first consider a case in which the \ac{manet} topology remains constant, and evaluate Unfolded PGDNet when given access to the true \ac{csi}. Our aim is to assert the ability of our unfolded design to cope with the non-convexity (\ref{itm:nonCvx}) and the need for rapid tuning (\ref{itm:rapid}). 
Fig,~\ref{fig:PGDNet vs PGD iters M=N=2}.
reports the min-rate of an Unfolded PGDNet with a single $E=1$ model compared with classic \ac{pgd} versus the iteration number, averaged over all $200$ channels, for a noise level of $0$ dB for all links.
We observe in Fig.~\ref{fig:PGDNet vs PGD iters M=N=2} that our learn-to-optimize method allows PGDNet to systematically approach the grid capacity with approximately $82 \times $ fewer iterations compared to conventional \ac{pgd} (which requires on average about $3300$ iterations to converge), demonstrating its ability to cope with \ref{itm:rapid}. 
 
\begin{figure}
    \centering
    \includegraphics[width=1\columnwidth]{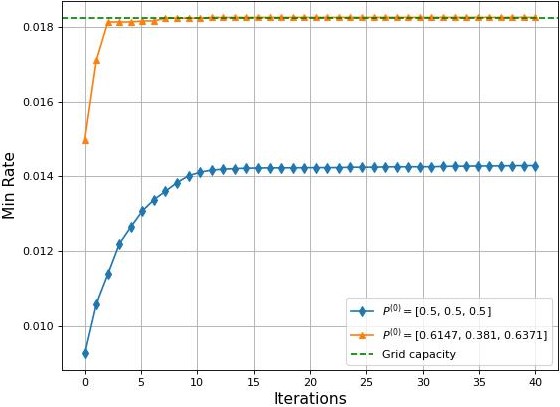}
   \vspace{-0.2cm}
    \caption{Min-rate per PGD iteration for a single channel realization, $1\times 2\times 2$ static \ac{manet}.} 
    \label{fig:ensemble channels}
\end{figure}

It is noted though that PGDNet with $E=1$ is still within some margin from the grid capacity due to the non-convex nature of the optimization problem \eqref{eqn: best power allocation}. This arises as for some of the channel realizations, starting from $\myMat{P}^{(0)}_1(t)$ does not allow to reach the grid capacity. However, when using an alternative initialization, Unfolded PGDNet  can rapidly achieve the grid capacity when starting from a different initial guess. This behavior is empirically demonstrated in Fig.~\ref{fig:ensemble channels}, confirming the need for our ensemble-based design and its ability to alleviate \ref{itm:nonCvx}.

\begin{figure}
    \centering
    \includegraphics[width=\columnwidth]{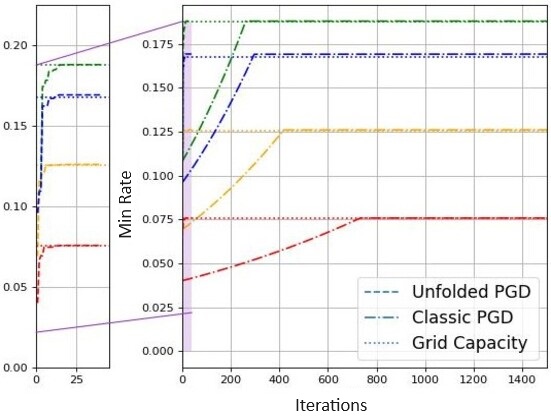}
   \vspace{-0.4cm}
    \caption{Min-rate vs. iteration for four different channel realizations, $1\times 2\times 2$ static \ac{manet}.} 
    \label{fig:PGDNet vs PGD capacity M=N=2}
   \vspace{-0.2cm}
\end{figure}

Figs.~\ref{fig:PGDNet vs PGD iters M=N=2} and \ref{fig:ensemble channels} indicate the validity of the main ingredients of Unfolded PGDNet, i.e., unfolding for rapid operation and ensemble for coping with non-convexity. We thus proceed by evaluating the overall Unfolded PGDNet using an ensemble of $E = 6$ models, where the initial points $\myMat{P}^{(0)}_2(t),\ldots,\myMat{P}^{(0)}_E(t)$ are randomized uniformly over $\mySet{P}(t)$.
The achieved rate versus the number of iterations for four random channel realizations are reported in Fig.~\ref{fig:PGDNet vs PGD capacity M=N=2}.
This figure demonstrates the robustness of the learned optimizer:  for each  of the randomly chosen channels, Unfolded PGDNet converged to the grid capacity much faster than the classic \ac{pgd}.
The reduction in latency achieved by Unfolded PGDNet here is of a factor of approximately $13.8\times$ ($K\times E=240$ iterations versus $3317$ required on average for \ac{pgd}).

We proceed by evaluating the rate achieved after $K$ iterations of Unfolded PGDNet, compared to a data-driven \ac{gnn}, which is not based on a principled optimizer. Fig.~\ref{fig:PGDNet vs PGD vs GNN full csi, (T, M, N) = (1, 2, 2)}
demonstrates that the PGDNet achieves the (mean) grid capacity in various SNR regimes where the \ac{gnn} fails.
In that topology, the classic \ac{pgd} also achieves grid capacity, but it takes $2000$ iterations, as opposed to its learned Unfolded PGDNet which achieves this performance within fixed $40$ iterations. 

\begin{figure}
    \centering
    \includegraphics[width=\columnwidth]{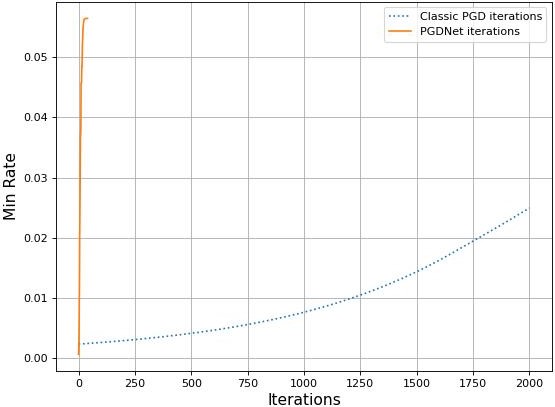}
   \vspace{-0.2cm}
    \caption{Min-rate vs. iteration averaged over $200$ channels   with full CSI, $1\times 3\times 3$ static \ac{manet}.} 
    \label{fig:PGDNet vs PGD iters M=N=3}
\end{figure}

The ability of Unfolded PGDNet to outperform the data-driven \ac{gnn} is preserved also for the larger $1\times 3 \times 3$ \ac{manet}, as reported in Fig.~\ref{fig:PGDNet vs PGD vs GNN full csi, (T, M, N) = (1, 3, 3)}. For such larger settings, computing the (mean) grid capacity becomes computationally prohibitive. We thus compare the min-rate achieved by three optimizers - classic \ac{pgd} with 2000 iterations, \ac{gnn}, and PGDNet with $K=40$. For every \ac{snr} regime, we observe in Fig.~\ref{fig:PGDNet vs PGD vs GNN full csi, (T, M, N) = (1, 3, 3)} that the proposed PGDNet consistently performs the best. It is emphasized that for such larger \acp{manet}, classic \ac{pgd} converges slower compared to smaller topologies. This behavior is observed  in Fig.~\ref{fig:PGDNet vs PGD iters M=N=3}, where the gaps between the learned unfolded algorithm and the fixed step-size version are notably more dominant compared to the smaller \ac{manet} reported in Fig.~\ref{fig:PGDNet vs PGD iters M=N=2}.

\begin{figure}
    \centering
    \includegraphics[width=\columnwidth]{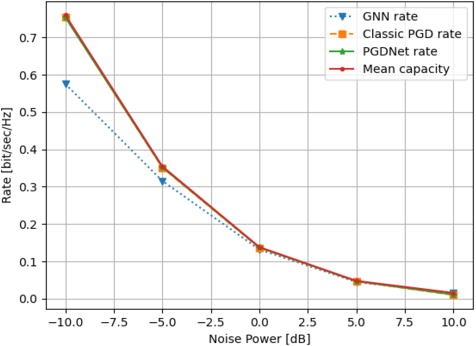}
   \vspace{-0.2cm}
    \caption{Min-rate vs. noise level  with full \ac{csi}, $1\times 2\times 2$ static \ac{manet}.} 
    \label{fig:PGDNet vs PGD vs GNN full csi, (T, M, N) = (1, 2, 2)}
\end{figure}

\begin{figure}
    \centering
    \includegraphics[width=\columnwidth]{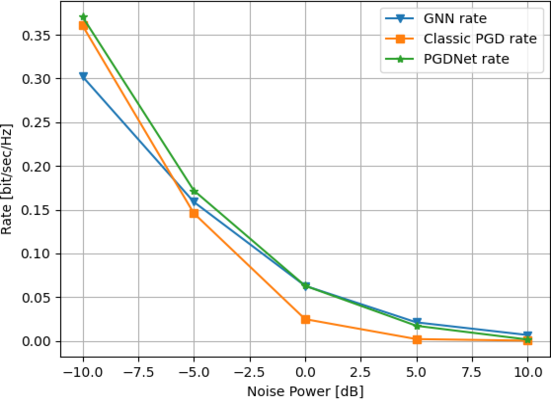}
   \vspace{-0.2cm}
    \caption{Min-rate vs. noise level with full CSI, $1\times 3\times 3$ static \ac{manet}.} 
    \label{fig:PGDNet vs PGD vs GNN full csi, (T, M, N) = (1, 3, 3)}
\end{figure}

\subsection{Static MANETs with Noisy \ac{csi}}
\label{subsec: static manets noisy}
We proceed by extending our experimental to the likely setting where there is no access to full \ac{csi} data, but instead one must use noisy pilots. We set the number of pilots to be the minimum allowed (i.e., $T= \max_{b,t}M_b(t)$). As detailed in Subsection~\ref{subsec:estData}, we employ \ac{lmmse} estimation to recover the noisy channel estimates as input features to our optimizers.
The training procedure uses estimated channels for the inputs during training, while the actual channel realizations are used for computing the loss during training via \eqref{eqn:trainLossNoisy}. 

\begin{figure}
    \centering
    \includegraphics[width=\columnwidth]{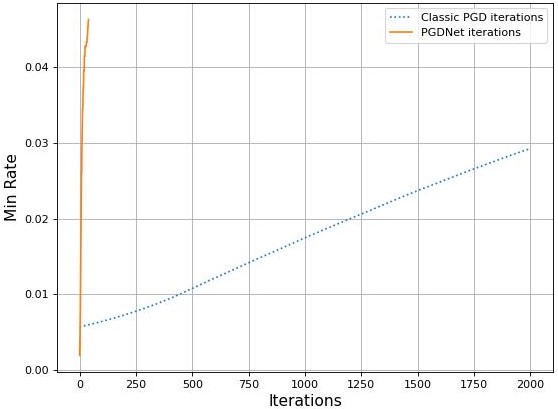}
   \vspace{-0.2cm}
    \caption{Min-rate vs. PGD iteration averaged over $200$ channels  with noisy CSI, $1\times 3\times 3$ static \ac{manet}.} 
    \label{fig:PGDNet vs PGD iters M=N=3 noisy model full csi}
\end{figure}

In Fig.~\ref{fig:PGDNet vs PGD iters M=N=3 noisy model full csi}, we evaluate the rate versus iterations for a noise level of $0$ dB.
We observe in Fig.~\ref{fig:PGDNet vs PGD iters M=N=3 noisy model full csi}  that the achieved min-rate is close to the one that was achieved using full \ac{csi} in Fig.~\ref{fig:PGDNet vs PGD iters M=N=3}. This indicates that our training method allows Unfolded PGDNet to deal with noisy \ac{csi}, converging at a similar rate as that with \ac{csi}.
Similar findings are also noted when observing the min-rate achieved after learned optimization is concluded for different noise levels. 
Furthermore, we can see in Fig.~\ref{fig:PGDNet vs PGD vs GNN full csi noisy csi, (T, M, N) = (1, 3, 3)}
that Unfolded PGDNet performs the best among the three algorithms (classic \ac{pgd}, \ac{gnn}, and PGDNet) in this scenario as well, preserving the trend observed with full \ac{csi}.

\begin{figure}
    \centering
    \includegraphics[width=\columnwidth]{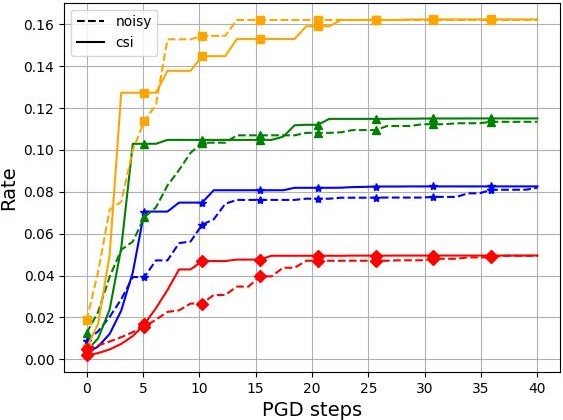}
   \vspace{-0.2cm}
    \caption{Min-rate vs. iteration for four  channel (each color represents a different realization), full \ac{csi}, $1\times 3 \times 3$ static \ac{manet}.} 
    \label{fig:four realizations 1-3-3 csi}
\end{figure}

\begin{figure}
    \centering
    \includegraphics[width=\columnwidth]{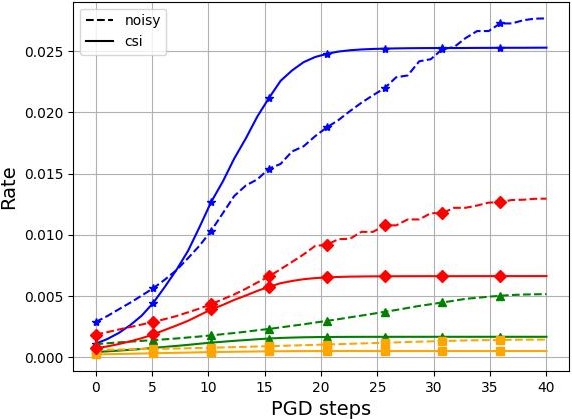}
   \vspace{-0.2cm}
    \caption{Min-rate vs. iteration for four channels (each color represents a different realization), noisy \ac{csi}, $1\times 3 \times 3$ static \ac{manet}.} 
    \label{fig:four realizations 1-3-3 noisy}
\end{figure}

While Unfolded PGDNet learns to cope with noisy \ac{csi}, its rate is still degraded compared to when processing the true \ac{csi}. To showcase this, we report in Fig.~\ref{fig:four realizations 1-3-3 csi} the rate versus iteration achieved by PGDNet trained with both full \ac{csi} and noisy \ac{csi} when applied to {\em full \ac{csi}}. For comparison, in Fig.~\ref{fig:four realizations 1-3-3 noisy} we report the rate versus iteration achieved when trained with both full \ac{csi} and noisy \ac{csi} and applied to {\em noisy \ac{csi}}. We observe in Fig.~\ref{fig:four realizations 1-3-3 csi} that there is indeed some minor degradation due to the processing of noisy channel estimates, yet the noisy training makes PGDNet robust and capable of reliably coping with noisy channel estimates. As expected, Unfolded PGDNet trained using full \ac{csi} consistently achieved improved min-data when indeed applied to full \ac{csi}, yet the performance gap compared to training with noisy \ac{csi} is quite minor.
However, in Fig.~\ref{fig:four realizations 1-3-3 noisy} we observe that noisy-aware training leads to improved robustness. In particular, Unfolded PGDNet trained in a noisy \ac{csi} aware manner consistently achieves non-negligible min-rate improvements compared to training using full \ac{csi} data, when applied to noisy \ac{csi} estimates. These results indicate on the usefulness of our noisy \ac{csi}-aware training, and its ability to robustify \ac{pgd}-based \ac{manet} optimization.

\begin{figure}
    \centering
    \includegraphics[width=\columnwidth]{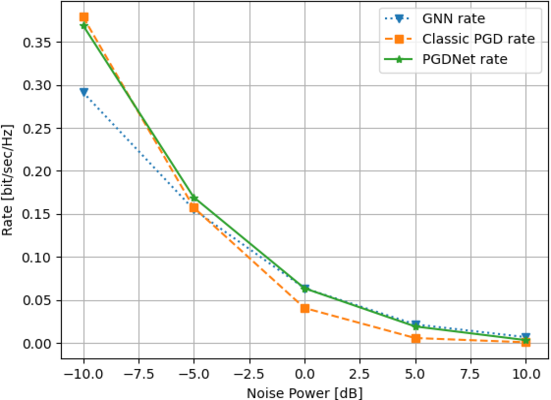}
   \vspace{-0.2cm}
    \caption{Min-rate vs. noise level with noisy CSI, $1\times 3\times 3$ static \ac{manet}.} 
    \label{fig:PGDNet vs PGD vs GNN full csi noisy csi, (T, M, N) = (1, 3, 3)}
\end{figure}

\begin{figure}
    \centering
    \includegraphics[width=\columnwidth]{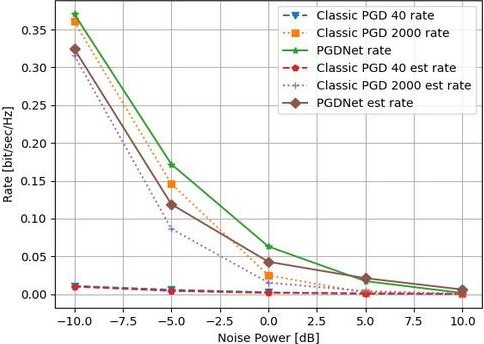}
   \vspace{-0.2cm}
    \caption{\ac{pgd} Vs. PGDNet applied to true \ac{csi} and noisy \ac{csi}, $1\times 3 \times 3$ static \ac{manet}} 
    \label{fig:PGd vs PGDNet 1-3-3 different settings}
\end{figure}

To further demonstrate the robustness of Unfolded PGDNet to noisy \ac{csi}, we compare in Fig.~\ref{fig:PGd vs PGDNet 1-3-3 different settings}
 classic \ac{pgd} with Unfolded PGDNet in both scenarios - access to full \ac{csi} data and estimated channels (coined {\em est} Fig.~\ref{fig:PGd vs PGDNet 1-3-3 different settings}) for different noise levels.
We compare three models - Unfolded PGDNet with $40$ iterations, classic \ac{pgd} with $40$ iterations, and classic \ac{pgd} with $2000$ iterations.
We can see in Fig.~\ref{fig:PGd vs PGDNet 1-3-3 different settings} that the gains of Unfolded PGDNet become more dominant in noisier settings, demonstrating the ability of the proposed methodology to notably facilitate optimizing superposition codes for \ac{noma} \acp{manet} in challenging channel conditions. 

\subsection{Time-Varying MANETs}
\label{subsec:time varying nets}
A key property of \acp{manet} is their ad-hoc dynamic nature. Accordingly, the network topology, which is typically multi-hop, may change randomly and rapidly over time (corresponding to the block-fading model). As discussed above, the training procedure is based on a given network topology, which is encapsulated in the channel realizations available in the data set \eqref{eqn:DataSet}. Nonetheless, as discussed in Subsection~\ref{subsec:PGDNet}, the learned parameters of PGDNet, i.e., the step-sizes s $\myVec{\mu}$, are independent of the topology of \ac{manet}. This allows Unfolded PGDNet to scale to different \ac{manet} sizes. 

To showcase the ability of PGDNet to be trained and applied in different topologies, representing time-varying \ac{manet}, we simulate a setting where we train on a given topology and use its learned $\myVec{\mu}$ to execute inference (calculating the best power allocation variables) on a different \ac{manet} topology. 
In particular, we train Unfolded PGDNet using channel realizations corresponding to a $1\times 2 \times 2$ \ac{manet}, and evaluate it on larger \acp{manet}, representing scenarios where additional devices are added to the network. Such settings are conceptually more challenging compared to removing devices, where one can model via a larger \ac{manet} while setting relevant links to have zero gain.

\begin{figure}
    \centering
    \includegraphics[width=\columnwidth]{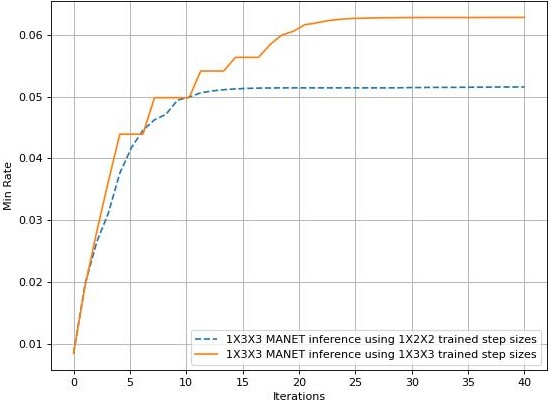}
   \vspace{-0.2cm}
    \caption
    {Inference a $M_1(t)=N(t)=3$ net using $M_1(t)=N(t)=2$ trained steps} 
    \label{fig: time-varying net 133}
\end{figure}
\begin{figure}
    \centering
    \includegraphics[width=\columnwidth]{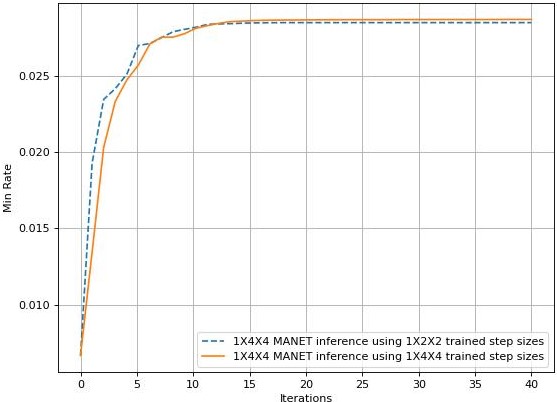}
   \vspace{-0.2cm}
    \caption
    {Inference a $M_1(t)=N(t)=4$ net using $M_1(t)=N(t)=2$ trained steps} 
    \label{fig: time-varying net 144}
\end{figure}
The scalability of Unfolded PGDNet is demonstrated in Figs.~\ref{fig: time-varying net 133}-\ref{fig: time-varying net 144}, where we compare the performance achieved by Unfolded PGDNet on topologies having $M_1(t)=3,N(t)=3$ and $M_1(t)=4,N_1(t)=4$,  respectively, whileused a trained $\myVec{\mu}$ of a $M_1(t)=2,N(t)=2$ network.
We observe that  Unfolded PGDNet trained on small \acp{manet} manages to reliably scale to larger \ac{manet}, achieving performance within a small gap to that achieved when trained for the larger topology. The min-rate gap is more notable for the $1\times 3 \times 3$ \ac{manet} compared to the $1\times 4 \times 4$, indicating that performance loss is not necessarily due to increasing the size of the \ac{manet}, but more likely due to changes in underlying symmetries (which are less profound when going from a $1\times 2 \times 2$ \ac{manet} to a $1\times 4 \times 4$ \ac{manet}).

\section{Conclusion}
\label{sec:Conclusions}
In this paper, we proposed Unfolded PGDNet for tuning superposition coding in two-hop \ac{noma} \acp{manet}. Unfolded PGDNet leverages data to operate reliably within a fixed and small number of iterations while being trainable in an unsupervised manner. We cope with the non-convexity of the problem by ensemble multiple models. Furthermore, we tackle the practical scenario of using pilots instead of relying on full \ac{csi} access successfully through \ac{lmmse} estimation.
We also manage to overcome the difficulty of the time-varying nature of \ac{manet}s by using already trained PGDNet that is empirically proven to be invariant in the manner of the \ac{manet} topology. 
Our numerical results show that Unfolded PGDNet rapidly sets superposition codes that approach the maximal achievable min-rate.


\begin{figure*}[ht!]    
\begin{align}
 \frac{\partial \log_2(f(\varphi_q(t)))}{\partial \varphi_q(t)}&=\left[\left(\frac{2|h_m^{(1)}|^2\varphi_q(t)(|h_m^{(1)}|^2\sum\limits_{\varphi_i(t) \in \Phi_n(t)} \varphi_i(t)^2+\sigma_1^2)}{(|h_m^{(1)}|^2\sum\limits_{\varphi_i(t) \in \Phi_n(t)} \varphi_i(t)^2+\sigma_1^2)^2}\right) - \left(\frac{2|h_m^{(1)}|^2\varphi_q(t)(|h_m^{(1)}|^2(\varphi_n(t)^2\sum\limits_{\varphi_i(t) \in \Phi_n(t)} \varphi_i(t)^2) +\sigma_1^2)}{(|h_m^{(1)}|^2\sum\limits_{\varphi_i(t) \in \Phi_n(t)} \varphi_i(t)^2+\sigma_1^2)^2}\right)\right] \notag \\ &\cdot \frac{1}{\ln{2}}  \left(\frac{|h_m^{(1)}|^2(\varphi_n(t)^2+\sum\limits_{\varphi_i(t) \in \Phi_n(t)} \varphi_i(t)^2)+\sigma_1^2}{|h_m^{(1)}|^2\sum\limits_{\varphi_i(t) \in \Phi_n(t)} \varphi_i(t)^2+\sigma_1^2}\right)^{-1} \notag \\&=
\frac{1}{\ln{2}} \frac{-2|h_m^{(1)}(t)|^4(\varphi_n(t))^2\varphi_q(t)}{|h_m^{(1)}(t)|^2\sum\limits_{i \in \mySet{N}(t)/n: \varphi_i(t) \geq \varphi_n(t) }(\varphi_i(t))^2 + \sigma_{1}^2} \frac{1}{|h_m^{(1)}(t)|^2\sum\limits_{i \in \mySet{N}(t)/n: \varphi_i(t) > \varphi_n(t) }(\varphi_i(t))^2 + \sigma_{1}^2}.
\label{eqn:RateGradProof}
\end{align}
\end{figure*}


	\vspace{-0.2cm}
	\begin{appendix}
		\numberwithin{equation}{subsection}	

		\subsection{Proof of Lemma \ref{lem:RateGrad}}
		\label{app:Proof1}
  
To prove \eqref{eqn:RateGrad}, recall that for any two functions $f_1(x), f_2(x)$ it holds that $\frac{d}{dx}\min(f_1(x),f_2(x)) = \frac{d}{dx}f_1(x)$ for every $x$ such that $f_1(x)<f_2(x)$. Consequently, computing the objective gradients boils down to \eqref{eqn:RateGrad}. Therefore, in order to prove the lemma, we show how  \eqref{eqn:RateGrad2} is obtained by taking the gradients of \eqref{eqn: bc rate} and \eqref{eqn: ma gain}.

Here, we derive the complex gradients of $R(\cdot)$ with respect to $\myMat{P}(t)$.  To that aim, we  examine the case 
where $\tilde{b} = 1$ and $R_{\tilde{m},n}^{(\tilde{b})}(t) < R_{l_{B},n}^{(B)}(t)$, and compute the partial derivative $\frac{\partial R_n(t)}{\partial \varphi_q(t)}$ where $\varphi_q(t) < \varphi_n(t)$. The gradients for the remaining hops are obtained using a similar derivation. 

 According to ~\eqref{eqn: bc rate}, $R_{m,n}^{{(1)}}(t) = \log_2(1 + \frac{|h_{m}^{(1)}(t)|^2 \varphi_n(t)^2}{|h_{m}^{(1)}(t)|^2 \sum\limits_{\varphi_i(t) \in \Phi_n(t)} \varphi_i(t)^2 + \sigma_{1}^2})$,
and $\frac{\partial \log_2(f(x))}{\partial x} = \frac{1}{\ln{2}} \frac{f'(x)}{f(x)}$.
Under this case, we get that the argument of the log expression (which we denote by $f(\cdot)$) satisfies 
\begin{align*}
f(\varphi_q(t)) &= 1 + (\frac{|h_{m}^{(1)}(t)|^2 \varphi_n(t)^2}{|h_{m}^{(1)}(t)|^2 \sum\limits_{\varphi_i(t) \in \Phi_n(t)} \varphi_i(t)^2 + \sigma_{1}^2})  \\&=
\frac{|h_{m}^{(1)}(t)|^2( \varphi_n(t)^2 +\sum\limits_{\varphi_i(t) \in \Phi_n(t)} \varphi_i(t)^2) + \sigma_1^2}  {|h_{m}^{(1)}(t)|^2 \sum\limits_{\varphi_i(t) \in \Phi_n(t)} \varphi_i(t)^2 + \sigma_{1}^2}.
\end{align*}
By taking the derivative of this equation with respect to $\varphi_q(t)$, one obtains the formulation in theorem ~\ref{lem:RateGrad} (and specifically of \eqref{eqn:RateGradRe}), as shown in \eqref{eqn:RateGradProof} (derived for  $\varphi_q(t) \in \Phi_n(t)$).
The derivation for the remaining gradients is obtained similarly, thus concluding the proof. \qed

\end{appendix}

\bibliographystyle{IEEEtran}

\bibliography{IEEEabrv,refs}

\end{document}